\documentclass[twocolumn,osajnl]{revtex4}

\pdfoutput=1

\usepackage{graphicx}
\usepackage{dcolumn}
\usepackage{bm}
\usepackage{color}

\usepackage{latexsym}
\usepackage{amsmath}
\usepackage{amsbsy} \usepackage{amsfonts}
\usepackage{amssymb} \usepackage{amscd} \usepackage{t1enc}
\usepackage{fleqn}
\usepackage{epsfig}

\usepackage{latexsym}
\usepackage{amsbsy}
\usepackage{amsfonts}
\usepackage{amssymb}
\usepackage{amscd}
\usepackage{amsmath}
\usepackage{rotating}

\usepackage{cases}
\usepackage{pst-all}

\usepackage[utf8]{inputenc} 
\usepackage[T1]{fontenc}

\renewcommand{\thesection}{\arabic{section}}

\def\p@subsection{}
\def\p@subsection     {\thesection\,}

\def\cha{\kern .6em{\sqcup \kern -1.12em \sqcup}\kern .6em}

\def\E{{\rm e}}
\def\I{{\rm i}}
\def\vec{\boldsymbol}
\def\q{\rm} %POUR ECRIRE LES QUATERNIONS DE BASE EN ROMAIN
\def\rap{\!\! } %POUR RAPPROCHER LES MEMBRES D'UNE EQUATION DANS eqnarray
\def\its{\it}

\def\isomor{\simeq}
\def\cal{\mathcal}

\def\CD{\hbox{
$\bigcirc {\kern-.75em { 
\hbox{$\buildrel > \over {\buildrel \over{}}$}}}$}}
\def\CG{\hbox{
$\bigcirc {\kern-.75em { 
\hbox{$\buildrel < \over {\buildrel \over{}}$}}}$}}
\mathchardef\Gamma="0100
\mathchardef\Delta="0101
\mathchardef\Theta="0102
\mathchardef\Lambda="0103
\mathchardef\Xi="0104
\mathchardef\Pi="0105
\mathchardef\Sigma="0106
\mathchardef\Upsilon="0107
\mathchardef\Phi="0108
\mathchardef\Psi="0109
\mathchardef\Omega="010A
{\bf}{\sl}
\newtheorem{proposition}{Proposition}[section]{\bf}{\its}
{\bf}{\rm}

\begin{document}

  \title{The twin paradox and time dilation  have equivalents  \\
    in polarization optics}

  \author{\bf Pierre Pellat-Finet}

\affiliation{Universit\'e Bretagne Sud,  UMR  CNRS  6205, LMBA, \\ F-56000 Vannes, France \\ pierre.pellat-finet@univ-ubs.fr}

\begin{abstract} {{\bf Abstract.}
  The phenomena known as the twin-paradox and time dilation, which are familiar effects in the special theory of relativity, have analogous counterparts in polarization optics. To show that, we present the concept of proper irradiance for a partially polarized lightwave. The analogous effect of the twin paradox is that the proper irradiance of the incoherent addition of two partially polarized lightwaves exceeds the  sum of the proper irradiances of each individual lightwave. This effect does not pose a paradox in optics. The analog  of time dilation is the increase in irradiance experienced by a lightwave as it propagates through a pure dichroic device.}
  %\end{abstract}


\vskip 1cm

\end{abstract}

\maketitle

%**********************************
\section{Introduction}\label{sect1}
%**********************************

Both the special theory of relativity and polarization optics are based on the same geometry, specifically that of the Minkowski space and  proper Lorentz rotations \cite{PPF1,PPF2}. 
As a result, certain relativistic phenomena have correspon\-ding manifestations in polarization optics, as they stem from underlying geometrical properties.
For instance, Lorentz boosts in rela\-ti\-vity and pure dichroics in polarization optics represent physical expressions of hyperbolic rotations in the abstract Minkowski space, leading to analogous effects in both domains. %realms.

The triangle inequality, which exhibits a particular form in  Minkowski space \cite{PPF2,Man},  gives rise to the well-known twin paradox in relativity \cite{PPF2,Mol,Rin,Sch}. In the present article,  we explore the implications of this inequality on the proper irradiance of a partially polarized lightwave, a concept equivalent to the proper time in relativity.
Our findings reflect the twin paradox in relativity within the context of polarization optics.
Those results align with established characteristics of partially polarized lightwaves and do not introduce any paradoxical elements in polarization optics.

Additionally, we interpret the increase in irradiance under a pure dichroic as analogous, in polarization optics, to time dilation in the realm of relativity.

%**************************************************
\section{Geometric analysis on Minkowski space}
%**************************************************

\subsection{Minkowski space}

We represent Minkowski's 4--vectors by minquats \cite{PPF1,PPF2,Syn}. (The term ``minquat,'' due to Synge \cite{Syn}, is an abbreviation for ``Minkowskian quaternion.'')
If $\vec x=(x_0,x_1,x_2,x_3)$ is a 4-vector in Minkowski space, the $x_\mu$'s ($\mu =0,1,2,3$) being real numbers, the associated  minquat is the complex quaternion $x$ such that
\begin{equation}
  x=x_0{\q e}_0+\I (x_1{\q e}_1+x_2{\q e}_2+x_3{\q e}_3)\,,\end{equation}
where ${\q e}_0$, ${\q e}_1$, ${\q e}_2$ and ${\q e}_3$ are unit quaternions, with
\begin{equation} ({\q e}_1)^2=({\q e}_2)^2=({\q e}_3)^2=-({\q e}_0)^2=-\,{\q e}_0\,,\end{equation}
\begin{equation}
  {\q e}_0{\q e}_j={\q e}_j={\q e}_j{\q e}_0\,,\hskip .2cm j=1,2,3,\end{equation}
\begin{equation}
  {\q e}_1{\q e}_2={\q e}_3=-\,{\q e}_2{\q e}_1\,.
\end{equation}

The Minkowski space is $({\mathbb R}^4,Q)$ where $Q$ denotes the Lorentz quadratic form, defined for every 4--vector $\vec x$ by
\begin{eqnarray}
  Q(\vec x)\rap &=&\rap Q(x_0,x_1,x_2,x_3)\nonumber \\
  \rap &=&\rap (x_0)^2-(x_1)^2-(x_2)^2-(x_3)^2\,.\end{eqnarray}

We denote as ${\mathbb M}$ the (real) vector space of minquats, a basis of which is $\{ {\q e}_0, \I\,{\q e}_1, \I\,{\q e}_2, \I\,{\q e}_3\}$. It is endowed with the quaternionic norm $N$, such that for every complex quaternion $q=q_0{\q e}_0+ q_1{\q e}_1+q_2{\q e}_2+q_3{\q e}_3$ (the $q_\mu$'s are complex numbers)
\begin{equation}N(q)=(q_0)^2+(q_1)^2+(q_2)^2+(q_3)^2\,.\end{equation}
If $x$ is a minquat, then
\begin{eqnarray}
  N(x)\rap &=& \rap N[x_0{\q e}_0+\I (x_1{\q e}_1+x_2{\q e}_2+x_3{\q e}_3)]\nonumber \\
  &=&\rap (x_0)^2-(x_1)^2-(x_2)^2-(x_3)^2\,.\end{eqnarray}
We conclude that the Minkowski space and the minquat space are quadratically isomorphic:
$({\mathbb M},N)\isomor({\mathbb R}^4,Q)$.

We say that
\begin{itemize}
\item[$\bullet$] $x$ is a spacelike minquat if $N(x)<0$;
\item[$\bullet$] $x$ is a null (or isotropic) minquat if $N(x)=0$;
\item[$\bullet$] $x$ is a scalarlike (or timelike) minquat if $N(x)>0$.
\end{itemize}
Correspondingly a 4--vector $\vec x$ may be spacelike, null or timelike, if $Q(\vec x)$ is strictly negative, zero or strictly positive.

\medskip
\noindent {\its Remark.} ``Scalarlike'' should not be confused with ``scalar''. A scalar quaternion takes the form $q=q_0\,{\q e}_0$, % ($q_0\in {\mathbb C}$),
whereas $x$ is a scalarlike minquat if its scalar component $x_0\,{\q e}_0$
is the dominant factor in $N(x)$, that is, if  $(x_0)^2> (x_1)^2+(x_2)^2+(x_3)^2$.

\subsection{Triangle inequality}%***********************************************

The triangle inequality in Minkowski space takes a form reverse to its form in an Euclidean space \cite{PPF2,Man}. If $\vec x$ and $\vec y$ are two 4--vectors belonging to Minkowski space, independently null or timelike, then (triangle inequality)
  \begin{equation}
    \sqrt{Q(\vec x+\vec y)}\ge \sqrt{Q(\vec x)}+\sqrt{Q(\vec y)} \,.\end{equation}
  The equality holds if, and only if, 4--vectors $\vec x$ and $\vec y$ are collinear.
  
  By isomorphism, the triangle inequality holds in the minquat space ${\mathbb M}$ under the form
    \begin{equation}
      \sqrt{N(x+y)}\ge \sqrt{N(x)}+\sqrt{N(y)} \,,\end{equation}
    where $x$ and $y$ are two independently null or scalarlike (timelike) minquats.

%***************************************
\section{The twin paradox in relativity}
%***************************************

\subsection{Basic elements of relativity \cite{Mol,Rin,Sch}}
In the special theory of relativity an event, denoted as ${\cal X}$, is represented by a 4--vector in  Minkowski space, say $\vec x$,  with coordinates $x_\mu$'s referring to an inertial (or  Galilean) frame (the $x_\mu$'s are real numbers). We set $x_0={\rm c}t$, where ${\rm c}$ denotes the speed of light (in a vacuum) and $t$ the time, in the considered frame. The $x_j$'s ($j=1,2,3$) are spatial coordinates along three orthogonal directions (also denoted as $x_j$).

If ${\cal R}$ and ${\cal R}'$ denote two inertial frames in relative uniform translatory  motion, the coordinates of an event in ${\cal R}$ and in ${\cal R}'$ are connected by a hyperbolic rotation \cite{PPF1,PPF2,Syn}, provided that the axis $x_j$ is parallel to the axis $x'_j$, and that $x'_0=0$ if, and only if, $x_0=0$. The event ${\cal X}$ is represented by the minquat $x$ in ${\cal R}$ and $x'$ in ${\cal R}'$, with $N(x)=N(x')$.

A spacetime interval between two events ${\cal X}$ and ${\cal Y}$ is also a 4--vector, represented 
by the minquat $y-x$ in ${\cal R}$ (coordinates $y_\mu -x_\mu$, $\mu =0,1,2,3$), and  $y'-x'$  in ${\cal R}'$ (coordinates $y'_\mu -x'_\mu$), with $N(y-x)=N(y'-x')$.

In the following we will denote a spacetime interval by $\Delta {\cal X}$, represented by the minquat $\Delta x$ in ${\cal R}$ (coordinates $\Delta x_\mu$), and by $\Delta x'$ in ${\cal R}'$ (coordinates $\Delta x'_\mu$). We have $N(\Delta x)=N(\Delta x')$.

We employ the notion of observer \cite{Rin,Sch},  attached to an inertial frame.
The spatial coordinates $x_j$ ($j=1,2,3$) of an observer ${\cal O}$ in an inertial frame, say ${\cal R}$,  may be regarded  as continuous functions of time $t$, or of para\-me\-ter  $x_0={\rm c}t$. 
We write $\Delta x_j(x_0)= x_j(x_0)-x_j(0)$, where $x_0=0$ is the origin of time in ${\cal R}$. The observer ${\cal O}$ is at rest in ${\cal R}$ if $\Delta x_j(x_0)=0$ ($j=1,2,3$) for every $x_0$ (${\cal R}$ is thus the frame to which ${\cal O}$ is attached). Therefore ${\cal O}$  is represented in ${\cal R}$ by the minquat $\Delta x={\q e}_0\Delta x_0$, and $\Delta x_0$ is called the proper time of ${\cal O}$. In another inertial frame, say  ${\cal R}'$, the observer ${\cal O}$ is represented 
by the minquat $\Delta x'={\q e}_0\Delta x'_0+\I({\q e}_1\Delta x'_1+{\q e}_2\Delta x'_2+{\q e}_3\Delta x'_3)$,
and  we have $N(\Delta x)=N(\Delta x')$, that is
\begin{equation}(\Delta x_0)^2=(\Delta x'_0)^2-(\Delta x'_1)^2-(\Delta x'_2)^2-(\Delta x'_3)^2.\;\,\end{equation}

\subsection{Twin paradox (timelike vectors)}%****************************************************************

\begin{figure}[b]%$$$$$$$$$$$$$$$$$$$$$$$$
\begin{center}
  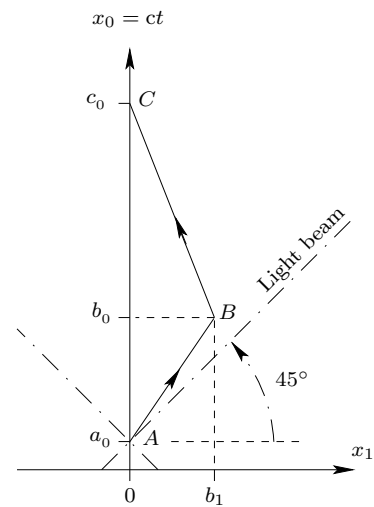
  \caption{Twin paradox in relativity. The proper time of the space-time interval $AC$ is greater that the sum of the individual proper times of $AB$ and $BC$. Dashed-dotted lines represent the trace of the light cone at $A$. The diagram is applicable in both Minkowski space and  minquat space ${\mathbb M}$, with  the latter involving basis vectors ${\q e}_0$ ($x_0$--axis) and $\I{\q e}_1$ ($x_1$--axis). \label{fig1}}
\end{center}
\vskip -.4cm
\end{figure}

An observer ${\cal O}$ stays at point $(x_1,x_2,x_3)=(0,0,0)$, in an inertial frame ${\cal R}$, while another observer ${\cal O}'$ travels from abscissa $x_1=0$ to  abscissa $x_1=b_1\ne 0$, on the $x_1$--axis, and then travels back to abscissa $x_1=0$.  The travel diagram is shown on Fig.\ \ref{fig1}: the observer ${\cal O}'$  travels from the point  $A=(a_0,0,0,0)$ to $B=(b_0,b_1,0,0)$ and then from $B$ to $C=(c_0,0,0,0)$. 
We have $c_0>b_0>a_0$.
Both motions of ${\cal O}'$ are accomplished in uniform translation with respect to ${\cal R}$. An inertial frame ${\cal R}'$ is attached to ${\cal O}'$, when travelling from $A$ to $B$, and an inertial frame ${\cal R}''$, when travelling from $B$ to $C$.

%$$$$$$$$$$$$$$$$$$$$$$$$$$$$$

The ``travel'' in time of ${\cal O}$ from $A$ to $C$ is represented in ${\cal R}$ by the minquat
\begin{equation}
  \Delta x=(c_0-a_0)\,{\q e}_0\,,\end{equation}
and the corresponding proper time is
\begin{equation}
  \Delta t_x= {1\over {\rm c}}\sqrt{N(\Delta x)}={1\over {\rm c}}\sqrt{(c_0-a_0)^2}\,.\end{equation}

The travel from $A$ to $B$ is represented in ${\cal R}$ by the minquat
\begin{equation}
  \Delta y=(b_0-a_0)\,{\q e}_0+ \I b_1 {\q e}_1\,,\end{equation}
and since ${\cal O}'$ is at rest in ${\cal R}'$, the corresponding proper time %(measured in ${\cal R}'$) 
is
\begin{equation}
  \Delta t'_y= {1\over {\rm c}}\sqrt{N(\Delta y)}={1\over {\rm c}}\sqrt{(b_0-a_0)^2-(b_1)^2}\,.\end{equation}
The travel from $B$ to $C$ is represented by
\begin{equation}
  \Delta z=(c_0-b_0)\,{\q e}_0- \I b_1 {\q e}_1\,,\end{equation}
and the corresponding proper time
is
\begin{equation}
  \Delta t''_z= {1\over {\rm c}}\sqrt{N(\Delta z)}={1\over {\rm c}}\sqrt{(c_0-b_0)^2-(b_1)^2}\,.\end{equation}
Since $\Delta x=\Delta y+\Delta z$, and since $\Delta y$ and $\Delta z$ are not collinear (because $b_1\ne 0$), the triangle inequality leads to
\begin{equation}
  \sqrt{N(\Delta x)}> \sqrt{N(\Delta y)}+ \sqrt{N(\Delta z)}\nonumber\,,
\end{equation}
that is
\begin{equation}
  \Delta t_x>\Delta t'_y+\Delta t''_z\,,\label{eq17}\end{equation}
which means that the travel time-length is measured greater by ${\cal O}$ than by ${\cal O}'$.

The whole-travel time-length  measured by ${\cal O}$ (in ${\cal R}$) is longer than the travel-length measured by ${\cal O}'$ (on his own clock). If the two observers are twins, ${\cal O}$ is older than ${\cal O}'$ when they meet again. That constitutes a paradox accor\-ding  to our common ways of conceiving time.

To compare with the equivalent effect in polarization optics (Sect.\ \ref{sect4}), we note that the speed of ${\cal O}'$ when tra\-vel\-ling from $A$ to $B$ is
\begin{equation}
  v_y={\rm c}\,{b_1\over b_0-a_0}\,,\end{equation}
and the speed in the travel from $B$ to $C$ is
\begin{equation}
  v_z={\rm c}\,{-b_1\over c_0-b_0}\,.\end{equation}
We  assume $b_1>0$ and define $\beta_y= v_y/{\rm c}$ and $\beta_z=-v_z/{\rm c}$; we have $0<\beta_y<1$ and $0<\beta_z<1$. The travel from $A$ to $B$ may be represented by the minquat
\begin{eqnarray}
  Y= \Delta y \rap &=&\rap (b_0-a_0)({\q e}_0+\I\beta_y {\q e}_1)\nonumber \\
  &=&\rap Y_0({\q e}_0+\I\beta_y {\q e}_1)\,,\end{eqnarray}
and the travel from $B$ to $C$ by
\begin{eqnarray}
  Z=\Delta z \rap &=&\rap (c_0-b_0)({\q e}_0-\I\beta_z {\q e}_1)\nonumber \\
  \rap &=&\rap Z_0({\q e}_0-\I\beta_z {\q e}_1)\,,\end{eqnarray}
with $Y_0\beta_y=Z_0\beta_z$ ($Y_0$ and $Z_0$ are strictly positive, because $c_0>b_0>a_0$).

We have $X={\q e}_0\,\Delta t_{x}=Y+Z=(Y_0+Z_0)\,{\q e}_0$, and 
eventually Eq.\ (\ref{eq17}) also writes
\begin{equation}
  Y_0+Z_0>Y_0\sqrt{1-{\beta_y}^{\! 2}}+Z_0\sqrt{1-{\beta_z}^{\! 2}}\,.\label{eq22}\end{equation}

%********************************************************
\section{Equivalence in polarization optics}\label{sect4}
%********************************************************

\subsection{Norm and degree of polarization\label{sect41}}

The polarization state of a partially polarized lightwave is represented by a minquat of the form
\begin{equation}
  X=X_0{\q e}_0+\I (X_1{\q e}_1+X_2{\q e}_2+X_3{\q e}_3)\,,
  \end{equation}
where the $X_\mu$'s are the Stokes parameters of the lightwave (they are homogeneous to irradiances and proportional to the lightwave power; moreover $X_0>0$) \cite{PPF2,Sim,Ram}.

The minquat $X$ can be written 
$X=X_0 ({\q e}_0 +\I\rho_{_X} {\q e}_n)$, where  $\rho_{_X}$ is the degree of polarization \cite{PPF2,Sim} of the lightwave ($0\le \rho_{_X}\le 1$), and where  ${\q e}_n$ is a real unit pure quaternion (${\q e}_n=n_1{\q e}_1+n_2{\q e}_2+n_3{\q e}_3$, with $\rho_{_X}n_j=X_j/X_0$, $j=1,2,3$, if $\rho_{_X}\ne 0$, so that $(n_1)^2+ (n_2)^2+(n_3)^2=1$). The unpolarized component of the lightwave is $(1-\rho_{_X})X_0\,{\q e}_0$; its polarized component is $X_0\rho_{_X}({\q e}_0+\I \,{\q e}_n)$, and  ${\q e}_n$ represents its polarization state on the Poincar\'e sphere \cite{PPF2,Ram}.

The norm of $X$ is
\begin{equation}
  N(X)=(X_0)^2(1-{\rho_{_X}}^{\! 2})\,.\end{equation}

The minquat $X$ represents a physical state of polarization if, and only if, $N(X)\ge 0$. If $X$ is a null (or isotropic) minquat, i.e. if $N(X)=0$, the corresponding lightwave is completely polarized. If $X$ is a scalarlike (or timelike) minquat, i.e. if $N(X)>0$, the lightwave is partially polarized.

The minquat $X={\q e}_0+\I\,{\q e}_n$ represents a unitary completely polarized state of polarization and  $X_\perp={\q e}_0-\I\,{\q e}_n$ the unitary orthogonal state.  By abuse we say  ``polarization state $X$,'' and ``orthogonal polarization state $X_\perp$.''

\subsection{Pure dichroics}\label{sect42}

The polarization of a lightwave may be altered  when the lightwave propagates through certain  media or reflects off smooth surfaces.  Among the physical devices that act on light polarization, we mention birefringent media and pure dichroic ones  \cite{PPF2} (a pure dichroic has an isotropic absorption equal to 1, whereas a physical passive dichroic has an isotropic absorption factor smaller than 1).  Birefringents and pure dichroics generate a group, the group of pure ``dephasers'' (or pure phase-shifters), which can be shown to be isomorphic to the group SO$_+(1,3)$ of proper Lorentz rotations on Minkowski space \cite{PPF2}.

On the other hand, proper Lorentz rotations can be represented by  quaternions of the form
$u=\exp {\q e}_u\psi$, where $\psi$ is a complex number and ${\q e}_u$ a complex pure quaternion (${\q e}_u$ may be a unit quaternion or a null quaternion \cite{PPF2}).  The proper Lorentz rotation re\-pre\-sented by $u$ operates on ${\mathbb M}$ according to
$X\longmapsto u\,X\,\overline{u}^{\, *}$, where $\overline{u}^{\, *}$ denotes the Hamilton conjugate of the complex conjugate of $u$. The transformation preserves the norm of minquats:
$N(u\,X\,\overline{u}^{\, *})=N(X)$.
If $\psi$ is a purely imaginary number and if ${\q e}_u$ is a real pure unit quaternion (then denoted as ${\q e}_n$), the previous Lorentz rotation is a hyperbolic rotation.

If polarization states are represented by minquats (see Sect.\ \ref{sect4}.1), a  pure dichroic operates on ${\mathbb M}$ as a hyperbolic rotation \cite{PPF2} and is represented by a unit quaternion of the form $u=\exp (\I\,{\q e}_n\delta /2)$, where $\delta$ is the dichroism (a real number) and ${\q e}_n$ is a real unit pure quaternion, called the axis of the dichroic (by reference to the Poincar\'e sphere representation of polarization states).
The image of the polarization state $X$ through the dichroic is $X'$ given by
\begin{equation}
  X'=u\,X\,u=\exp \left(\I\,{\q e}_n{\delta\over 2}\right)\,X\,\exp \left(\I\,{\q e}_n{\delta\over 2}\right)\,,\label{eq25}\end{equation}
because $u=\overline{u}^{\, *}$ for a dichroic.

For actual derivations we use
\begin{equation}
  \exp {\I\,{\q e}_n\delta\over 2}={\q e}_0\cosh{\delta \over 2}+\I \,{\q e}_n\sinh{\delta \over 2}\,.
  \end{equation}
For example, if ${\q e}_n={\q e}_1$, Eq.\ (\ref{eq25}) gives
\begin{eqnarray}
  X'_0\rap &=&\rap X_0\,\cosh\delta +X_1\,\sinh\delta\,,\label{eq27a}\\
 X'_1 \rap &=&\rap X_0\,\sinh\delta +X_1\,\cosh\delta\,,\label{eq28a}\\
  X'_2 \rap &=&\rap X_2\,,\label{eq29a}\\
  X'_3 \rap &=&\rap X_3\,.\label{eq30a}\end{eqnarray}

If $u=\exp (\I\,{\q e}_n\delta /2)$, the minquats $X^+={\q e}_0+\I \,{\q e}_n$ and  $X^-={\q e}_0-\I\, {\q e}_n$ represent two orthogonal unit eigenstates of polarization. They are such that  $u\,X^+\,u=\E^\delta X^+$ and  $u\,X^-\,u=\E^{-\delta} X^-$,
which means that for $\delta >0$, the polarization state $X^+$ is amplified, whereas the state $X^-$ is mitigated.

In general, dichroics are ``passive'' devices, they cannot amplify the incident-lightwave power. To represent them, we introduce an isotropic absorption factor $\kappa$ ($\kappa>0$), such that $\kappa \exp \delta <1$ (for  $\delta >0$), so that a dichroic is represented by a quaternion of the form $\sqrt{\kappa}\,\exp (\I\,{\q e}_n\delta /2)$. It should be clear that a passive dichroic does not preserve the norm of minquats as pure dichroics do.

A  pure dichroic ($\kappa =1$) can be obtained with an active medium: an eigenstate is amplified, whereas the orthogonal state is mitigated. For instance, this can be achieved in a laser cavity.

\subsection{Proper irradiance}\label{sect43}%*********************************************

Let us consider the state $X=X_0({\q e}_0+\I\rho_{_X}\,{\q e}_n)$ and a pure dichroic of axis ${\q e}_n$ and dicroism $\delta$. The image of $X$ under the dichroic is
\begin{eqnarray}
  X'\rap &=&\rap\exp\left(\I\,{\q e}_n{\delta \over 2}\right)\,X\,\exp\left(\I\,{\q e}_n{\delta \over 2}\right)\nonumber \\
  &=&\rap  X'_0({\q e}_0+\I\rho_{_{X'}}{\q e}_n)\,,\end{eqnarray}
where
\begin{equation}
  X'_0=X_0(\cosh\delta +\rho_{_X}\sinh\delta )\,,\end{equation}
and
\begin{equation}
  \rho_{_{X'}}={\sinh\delta +\rho_{_X}\cosh\delta\over \cosh\delta +\rho_{_X}\sinh \delta}\,.\end{equation}

We assume that $X$ does not represent a completely polarized lightwave:  $0\le \rho_{_X}<1$ (i.e. the lightwave is strictly partially polarized and $X$ is a scalarlike minquat). Let us denote $\widetilde \delta$ the value of $\delta$ such that $\tanh \widetilde\delta =-\rho_{_X}$, for which $X'$ becomes $\widetilde X$. Then $\rho_{_{\widetilde X}}=0$, and $\widetilde X$ is a scalar quaternion: $\widetilde X=\widetilde X_0\,{\q e}_0$; it represents an unpolarized lightwave.

We remark that $\widetilde\delta <0$, which means that the state ${\q e}_0+\I\,{\q e}_n$ is the mitigated eigenstate of the dichroic (with eigenvalue $\exp \widetilde\delta <1$), whereas the state  ${\q e}_0-\I\,{\q e}_n$ is the amplified eigenstate (eigenvalue $\exp (-\widetilde\delta \,)>1$).

From $\tanh\widetilde\delta = -\rho_{_X}$, we deduce
\begin{equation}
  \widetilde X=\widetilde X_0\,{\q e}_0={X_0\over \cosh\widetilde\delta }\,{\q e}_0\,,\label{eq29}\end{equation}
and
\begin{eqnarray}
  (\widetilde X_0)^2={(X_0)^2\over \cosh^2\widetilde \delta}= N(\widetilde X)\rap &=&\rap N(X)\nonumber \\
  \rap &=&\rap (X_0)^2(1-{\rho_{_X}}^{\! 2})\,.\label{eq30}
\end{eqnarray}
(We have $N(\widetilde X)= N(X)$, because $X$ is transformed into $\widetilde X$ under a pure dichroic.)

We call $\widetilde X_0$ the proper irradiance of the lightwave represented by $X$ and we state the following proposition:
\begin{proposition} For every partially polarized lightwave whose degree of polarization is strictly smaller than $1$, there exists a pure dichroic that transforms it into an unpolarized lightwave. The irradiance of the  unpolarized lightwave is called the proper irradiance of the initial lightwave.
\end{proposition}

The proper irradiance  $\widetilde X_0$ of a lightwave represented by the scalarlike minquat $X$ has the following property. Let
 us consider an arbitrary pure dichroic that transforms $X$ into $X''$, with % The minquat $X''$ takes the form
$X''=X''_0({\q e}_0+\I\rho_{_{X''}}{\q e}_m)$. Then
\begin{eqnarray}
  (X''_0)^2(1-{\rho_{_{X''}}}^{\!\! 2})=N(X'') \rap &=&\rap N(X)\nonumber \\
  &=&\rap N(\widetilde X)=(\widetilde X_0)^2\,,
\end{eqnarray}
so that
\begin{equation}
  \widetilde X_0\le X''_0\,.\label{eq32a}
\end{equation}

Let  ${\frak D}_X$  be the set of all minquats $X''$ that are images of $X$ under all pure dichroics. Such a minquat $X''$ is given by $X''=\exp (\I\, {\q e}_\ell\delta /2)\,X\,\exp (\I\, {\q e}_\ell\delta /2)$, where $\delta$ is a real number and where ${\q e}_\ell$ is a real unit pure quaternion. It can also be written
$X''=X''_0({\q e}_0+\I\rho_{_{X''}}\,{\q e}_m)$ and then,
according to Eq.\ (\ref{eq32a})
\begin{equation}\widetilde X_0=\min\limits_{X''\in \,\frak D_X}\!\!X''_0\,.\label{eq33a}\end{equation}

Equation (\ref{eq33a}) delineates a characteristic property of the proper irradiance of a partially polarized lightwave: it repre\-sents the minimum achievable irradiance  when the lightwave traverses a pure dichroic.

\subsection{The equivalent of twin paradox}\label{sect44}%*******************************************

Let $Y=Y_0({\q e}_0+\I\rho_{_Y} \,{\q e}_n)$ and $Z=Z_0({\q e}_0-\I\rho_{_Z} \,{\q e}_n)$ represent two strictly partially polarized lightwaves ${\cal Y}$ and ${\cal Z}$ with $Y_0\rho_{_Y}=Z_0\rho_{_Z}\ne 0$. Lightwaves ${\cal Y}$ and ${\cal Z}$ have orthogonal polarized components. Let ${\cal X}$ be the incoherent addition \cite{Ram} (or superposition) of  ${\cal Y}$ and ${\cal Z}$. The polarization state of ${\cal X}$ is then represented by the minquat
\begin{equation}X=Y+Z= (Y_0+Z_0)\,{\q e}_0\,,\end{equation}
and corresponds to an unpolarized lightwave, whose degree of polarization is $0$. Since $X$, $Y$ and $Z$ are scalarlike minquats, not collinear, the triangle inequality gives
\begin{equation}
  \sqrt{N(X)}> \sqrt{N(Y)} +\sqrt{N(Z)}\,,\end{equation}
that is
\begin{equation}
  X_0=Y_0+Z_0>Y_0 \sqrt{1-{\rho_{_Y}}^{\! 2}}+Z_0 \sqrt{1-{\rho_{_Z}}^{\! 2}}\,.\label{eq34n}\end{equation}
Equation (\ref{eq34n}) is no more than Eq.\ (\ref{eq22}) in which $\beta_y$ and $\beta_z$ are replaced with $\rho_{_Y}$ and $\rho_{_Z}$.
Moreover, since $X$ represents an unpolarized lightwave, we have $\widetilde X_0=X_0$, and Eq.\ (\ref{eq30}), applied to $Y$ and $Z$, leads us to write Inequality (\ref{eq34n}) in the form
\begin{equation}
  \widetilde X_0>\widetilde Y_0+\widetilde Z_0\,,\label{eq37}\end{equation}
which means that the proper irradiance of $X$ is greater than the sum of the individual proper irradiances of $Y$ and $Z$. Equation (\ref{eq37}) is equivalent, in polarization optics,  to Eq.\ (\ref{eq17}) in relativity.

\subsection{Graphical equivalence}%*******************************
\begin{figure}[b]%$$$$$$$$$$$$$$$$$$$$$$$$
\begin{center}
  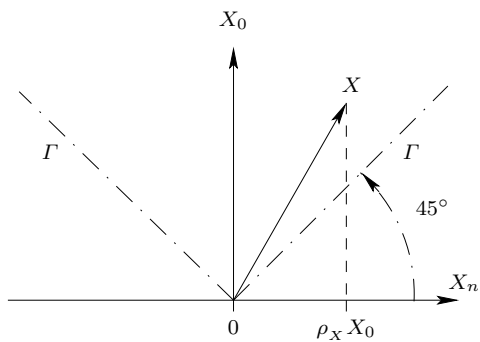
  \caption{Graphical representation of a polarization state in ${\mathbb M}$.
    The minquat $X=X_0({\q e}_0+\I\rho_{_X}{\q e}_n)=X_0\,{\q e}_0+\I X_n\,{\q e}_n$ lies in the plane spanned by ${\q e}_0$ ($X_0$--axis) and $\I\,{\q e}_n$ ($X_n$--axis). The dashed--dotted lines are the trace of the null-minquat half-cone $\Gamma$, whose surface represents the set of completely-polarized states. On the diagram, the min\-quat $X$ belongs to the interior  of $\Gamma$: it is then a scalarlike minquat and it represents a partially-polarized state.\label{fig2}}
\end{center}
\vskip -.4cm 
\end{figure}%$$$$$$$$$$$$$$$$$$$$$$$$$$$$

If the minquat $X$ represents a partially or completely polarized lightwave, then $X_0\ge 0$ and $N(X)\ge 0$. If $Y=\alpha X$, with $\alpha >0$, then $Y_0=\alpha X_0\ge 0$ and $N(Y)=\alpha^2N(X)\ge 0$, so that $Y$ also represents a partially or completely polarized lightwave. We conclude that minquats that represent lightwave polarization states form  a subset of ${\mathbb M}$ which is a half-cone $\Gamma$ and its interior. The half-cone equation is
\begin{equation}
  N(X)= 0\,,\hskip .5 cm X_0\ge 0\,.\end{equation}
The symmetry-axis of the cone is the straightline of equation $X_1=X_2=X_3=0$.

In polarization optics the half-cone $\Gamma$ is the equivalent of the future light-cone in relativity (Fig.\ \ref{fig1}).

The minquat $X=X_0({\q e}_0+\I\rho_{_X}{\q e_n})=X_0+\I X_n{\q e}_n$ represents a polarization state and lies in the plane $({\q e}_0,\I\, {\q e}_n)$ (included in ${\mathbb M}$), as shown in Fig.\ \ref{fig2}.

\begin{figure}[h]%$$$$$$$$$$$$$$$$$$$$$$$$
\begin{center}
  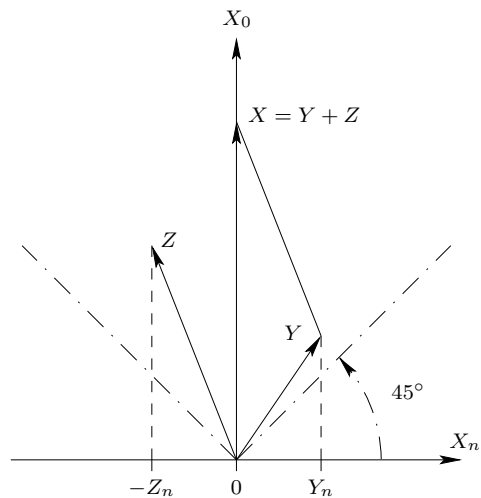
  \caption{Graphical equivalent representation of the twin paradox in polarization optics. Minquats $Y$ and $Z$ are such that $Y_n=Y_0\rho_{_Y}=Z_0\rho_{_Z}=Z_n$ (Sect.\ \ref{sect4}.4). The minquat $X$ represents the incoherent addition of the polarization states represented by $Y$ and $Z$.  The proper irradiance of $X$ is greater than the sum of the proper irradiances of $Y$ and $Z$. To be compared with Fig.\ \ref{fig1}.\label{fig3}}
  \end{center}
    \end{figure}

The incoherent addition of states $Y$ and $Z$ as in Sect.\ \ref{sect4}.4 is represented in Fig.\ \ref{fig3}. The comparison with Fig.~\ref{fig1} emphasizes the analogy with the twin paradox.

%$$$$$$$$$$$$$$$$$$$$$$$$$
 
\subsection{Conclusion (twin paradox for scalarlike minquats)}%**********************************************

We summarize the preceding findings into the following three equivalent statements, applicable to scalarlike vectors or minquats:
\begin{itemize}
\item[$\bullet$] {\its Minkowski-space geometry (triangle inequality).}
  The square root of the norm of the sum of two scalarlike (timelike) minquats %exceeds
  is greater than the sum of the square roots of the individual norms.
\item[$\bullet$] {\its Special Relativity (twin paradox).} 
  The proper time of the sum of two timelike 4-vectors in Minkowski spacetime is greater than the sum of the individual proper times.
\item[$\bullet$] {\its Polarization optics.} The proper irradiance of the incoherent addition %combination
  of two strictly partially polarized lightwaves is greater that the sum of the individual proper irradiances.
\end{itemize}

\subsection{Particular case. Expression with the degree of polarization}%************************************

We  assume
$Y=Y_0({\q e}_0+\I\rho\,{\q e}_m)$ and $Z=Z_0({\q e}_0+\I\rho\,{\q e}_n)$, with ${\q e}_m\ne {\q e}_n$ (and $\rho_{_Y}=\rho_{_Z}=\rho$). The incoherent superposition of $Y$ and $Z$ is
\begin{equation}
  X=Y+Z=(Y_0+Z_0)\,{\q e}_0+\I \rho \,(Y_0{\q e}_m+Z_0{\q e}_n),\;
\end{equation}
which writes
\begin{equation}
  X=(Y_0+Z_0)({\q e}_0+\I \rho_{_X}\,{\q e}_\ell)\,,\end{equation}
where ${\q e}_\ell$ is a real unit pure quaternion and $0\le \rho_{_X}<1$.

Since ${\q e}_m\ne {\q e}_n$ (which means that $Y$ and $Z$ are not collinear), the triangle inequality gives
\begin{equation}
  1-{\rho_{_X}}^{\! 2}> 1-\rho^2,\label{eq40}
\end{equation}
which is a special case of Eq.\  (\ref{eq37}), according to Eq.\ (\ref{eq30}). Equation (\ref{eq40}) leads to
\begin{equation}
  \rho_{_X}<\rho\,,\label{eq41}\end{equation}
that is,  the degree of polarization of the incoherent superposition of $Y$ and $Z$
is smaller than the degree of polarization of $Y$ and $Z$. Equation (\ref{eq41})
  may be regarded as a consequence of Eq.\ (\ref{eq37}), expressing the equivalent of twin paradox in optics in terms of the degree of polarization. (Similar examples are given in a recent publication \cite{PPF2}.)

More generally, the degree of polarization of the incoherent addition of two partially polarized lightwaves
is smaller than---or at most equal to---the degree of polarization of the indi\-vi\-dual lightwaves. That property of partially polarized lightwaves, analogous to the relativistic twin-paradox, is well known and does not constitutes any paradox in polarization optics.

%****************************************************************
\section{Time dilation and its equivalent in polarization optics}
%****************************************************************

\subsection{Time dilation in relativity}

Let us consider a spacetime interval represented by the minquat $\Delta x\ne 0$, in the inertial frame ${\cal R}$, and by $\Delta x'\ne 0$, in ${\cal R}'$. We assume ${\cal R}'$ in  translatory motion with velocity $v\,{\q e}_1$ with respect to ${\cal R}$ ($|v|<{\rm c}$). Minquats $\Delta x$ and $\Delta x'$ are linked by a hyperbolic rotation according to
\begin{equation}
  \Delta x'=\exp \left(\I\,{\q e}_1{\delta\over 2}\right)\,\Delta x\,\exp \left(\I\,{\q e}_1{\delta\over 2}\right)\,,\label{eq27}\end{equation}
where $\delta$ ($\delta\ne 0$) is the rapidity of the motion, such that
\begin{eqnarray}
  \cosh\delta \rap& =&\rap \gamma ={1\over \sqrt{1-\displaystyle{v^2\over {\rm c}^2}}}\,,\\
  \sinh \delta \rap& =&\rap -\gamma \beta =-\gamma {v\over {\rm c}}\,.\end{eqnarray}
Equation (\ref{eq27}) is equivalent to the following equations
\begin{eqnarray}
  \Delta x'_0\rap&=&\rap \Delta x_0\,\cosh\delta +\Delta x_1\,\sinh\delta\,,\label{eq45}\\
  \Delta x'_1 \rap&=&\rap \Delta x_0\,\sinh\delta +\Delta x_1\,\cosh\delta\,,\label{eq46}\\
  \Delta x'_2\rap&=&\rap \Delta x_2\,,\\
  \Delta x'_3\rap&=&\rap\Delta x_3\,.\end{eqnarray}

Now, let us assume the spacetime interval such that  $\Delta x'_j=0$ ($j=1,2,3$), which means that it is an interval between two events that take place at the same ``spatial'' position in ${\cal R}'$. Its proper time is equal to $\Delta x'_0$. For instance, the considered interval is a time interval given by a clock at rest in ${\cal R}'$. From  $\Delta x'_1=0$, we deduce $\Delta x_1=-\Delta x_0\tanh\delta$ and then, by Eq.\ (\ref{eq45}) 
\begin{equation}
  \Delta x_0=\Delta x'_0 \,\cosh\delta >\Delta x'_0\,.\label{eq47}\end{equation}
The time interval $\Delta x'_0$, which is the proper time of a clock at rest in ${\cal R}'$, is measured greater in ${\cal R}$. That is time dilation.

Time dilation is reciprocal. Let us assume, indeed, that $\Delta x_j=0$ ($j=1,2,3$). Thus $\Delta x_0$ is the proper time given by a clock at rest in ${\cal R}$. From Eq.\ (\ref{eq45}) we obtain
\begin{equation}
  \Delta x'_0= \Delta x_0\,\cosh\delta>\Delta x_0\,,\label{eq50}
\end{equation}
with $\delta$ as above. The proper time given by the clock at rest in ${\cal R}$ is measured greater in ${\cal R}'$.

\subsection{Time-dilation equivalent in polarization optics}%***********************************************
The equivalent of time dilation results from the representation of a pure dichroic by a hyperbolic rotation (see Sect.\ \ref{sect4}.2).
If $X$ represents the state of polarization of a lightwave incident on a dichroic medium of dichroism $\delta$ ($\delta >0$) and axis ${\q e}_1$ on the Poincar\'e  sphere, the emerging state of polarization is represented by the minquat $X'$, given by Eq.\ (\ref{eq25}) for ${\q e}_n={\q e}_1$, that is
\begin{equation}
  X'=\exp \left(\I\,{\q e}_1{\delta\over 2}\right)\,X\,\exp \left(\I\,{\q e}_1{\delta\over 2}\right)\,.\label{eq35}\end{equation}
Equation (\ref{eq35}) is no more than Eq.\ (\ref{eq27}) where the rapidity in replaced with the dichroism and spacetime intervals with polarization-state representative minquats.

Let $X=X_0({\q e}_0+\I\rho_{_X}{\q e}_1)$ represent a strictly partially polarized lightwave and let $\exp (\I\,{\q e}_1\widetilde \delta /2)$ represent the pure dichroic that transforms $X$ into an unpolarized lightwave $\widetilde X$ (see Sect.\ \ref{sect4}.3). The proper irradiance of the lightwave is $\widetilde X_0$, and according to Eq.\ (\ref{eq29})
\begin{equation}
  X_0=\widetilde X_0 \,\cos\widetilde \delta>\widetilde X_0\,,
  \end{equation}
which is Eq.\ (\ref{eq47}) where the rapidity $\delta$ is replaced with the dichroism $\widetilde \delta$, the proper time $\Delta x'_0$ with the proper irradiance $\widetilde X_0$, and $\Delta x_0$ with the irradiance $X_0$.

For the reciprocal effect,
let us assume that $X$ takes the form $X=X_0\,{\q e}_0$, and thus corresponds to an unpolarized lightwave. Then $X_0$ is also the proper irradiance of $X$, that is, $\widetilde X_0=X_0$.

According to Eq.\ (\ref{eq35}), the minquat $X'$ takes the form
\begin{eqnarray}
  X'\rap &=&\rap X_0\cosh\delta +\I X_0\,{\q e}_1\sinh\delta\nonumber \\
   &=&\rap X_0\cosh\delta \;({\q e}_0+\I\rho_{_{X'}}{\q e}_1)\nonumber \\
  &=&\rap X'_0\,({\q e}_0+\I\rho_{_{X'}}{\q e}_1)\,,
  \end{eqnarray}
where $\rho_{_{X'}}$ denotes the corresponding degree of polarization, and
\begin{equation}
  X'_0=X_0\cosh\delta=\widetilde X_0\cosh\delta >\widetilde X_0\,.\label{eq51}\end{equation}

For the analogy with time dilation, we note that Eq.\ (\ref{eq51}) is Eq. (\ref{eq50}) in which $\Delta x'_0$ (time interval) is replaced with $X'_0$ (irradiance), the proper time $\Delta x_0$ is replaced with the proper irradiance $\widetilde X_0$, and the rapidity with the dichroism.

Equation (\ref{eq51})   means that the irradiance of the emerging lightwave is greater than the proper irradiance of $X$.
Moreover, a dichroic is also a partial polarizer:  $X'$ is partially polarized, whereas $X$ is unpolarized. Indeed $\rho_{_{X'}}\!>0$, because $\delta >0$ : the degree of polarization of $X'$ is greater than that of $X$ (which is zero).
A part of the unpolarized component of the input state $X$ is transformed into the polarized component of the output state $X'$. To obtain $N(X')=N(X)$, the irradiance $X'_0$ has to be greater than $\widetilde X_0$.

Eventually we point out that Lorentz boosts in relativity and pure dichroics in polarization optics are physical manifestations of hyperbolic rotations in the abstract Minkowski space, resulting in comparable physical consequences, including time dilation and its equivalent in optics.

%*******************
\section{Conclusion}
%*******************
Although widely accepted by the scientific community,
certain relativistic effects like  time dilation or twin paradox are perceived counterintuitive, sparking debates over their interpretations.
Actually these effects stem from the geometry of Minkowski space; this one, however, is not the underlying vector space of the sole special theory of relati\-vi\-ty: it also constitutes a basic structure of polarization optics. The aforementioned relativistic effects have then their counterparts  in polarization optics and this area of physics affords to illustrate them, as we have  shown in this article. 
Polarization optics also offers a means to expe\-ri\-men\-tally test  geometrical properties of Minkowski space, while avoiding the paradoxes, whether real or perceived, linked to their interpretations in relativity.

\smallskip
\noindent {\bf Acknowledgement.}
The author gratefully acknowledges Prof. Jorge Mahecha Gomez (Universidad de Antioquia, Medell\'in, Colombia) for helpful comments and suggestions.

%**************************************
%BIBLIOGRAPHY**************************

%*************
\end{document}